\RequirePackage{lineno}
\documentclass[aps,prb,twocolumn,superscriptaddress]{revtex4}
\usepackage{graphicx}
\usepackage{epstopdf}
\usepackage{color}
\usepackage{lettrine}
\usepackage{type1cm}
\usepackage{amsmath}

\usepackage[font=sf,labelfont=bf,justification=raggedright]{caption}
\DeclareCaptionLabelSeparator{line}{ $|$ }
\begin{document}

\title{\textsf{Current-induced skyrmion generation and dynamics in symmetric bilayers}}

\author{A.~Hrabec}
\affiliation{Laboratoire de Physique des Solides, CNRS UMR 8502, Universit\'{e}s Paris-Sud et Paris-Saclay, 91405 Orsay Cedex, France}
\author{J.~Sampaio}
\affiliation{Laboratoire de Physique des Solides, CNRS UMR 8502, Universit\'{e}s Paris-Sud et Paris-Saclay, 91405 Orsay Cedex, France}
\author{M.~Belmeguenai}
\affiliation{LSPM (CNRS-UPR 3407), Universit\'{e} Paris 13, Sorbonne Paris Cit\'{e}, 99 avenue Jean-Baptiste Cl\'{e}ment, 93430 Villetaneuse, France}
\author{I.~Gross}
\affiliation{Laboratoire Aim\'{e} Cotton, CNRS, Universit\'{e} Paris-Sud, ENS Cachan, Universit\'{e} Paris-Saclay, 91405 Orsay Cedex, France}
\affiliation{Laboratoire Charles Coulomb, Universit\'{e} de Montpellier and CNRS UMR 5221, 34095 Montpellier, France}
\author{R.~Weil}
\affiliation{Laboratoire de Physique des Solides, CNRS UMR 8502, Universit\'{e}s Paris-Sud et Paris-Saclay, 91405 Orsay Cedex, France}
\author{S.M.~Ch\'{e}rif}
\affiliation{LSPM (CNRS-UPR 3407), Universit\'{e} Paris 13, Sorbonne Paris Cit\'{e}, 99 avenue Jean-Baptiste Cl\'{e}ment, 93430 Villetaneuse, France}
\author{A.~Stashkevich}
\affiliation{LSPM (CNRS-UPR 3407), Universit\'{e} Paris 13, Sorbonne Paris Cit\'{e}, 99 avenue Jean-Baptiste Cl\'{e}ment, 93430 Villetaneuse, France}
\affiliation{International Laboratory MultiferrLab, ITMO University, St. Petersburg 197101, Russia}
\author{V.~Jacques}
\affiliation{Laboratoire Charles Coulomb, Universit\'{e} de Montpellier and CNRS UMR 5221, 34095 Montpellier, France}
\author{A.~Thiaville}
\affiliation{Laboratoire de Physique des Solides, CNRS UMR 8502, Universit\'{e}s Paris-Sud et Paris-Saclay, 91405 Orsay Cedex, France}
\author{S.~Rohart}
\email[]{stanislas.rohart@u-psud.fr}
\affiliation{Laboratoire de Physique des Solides, CNRS UMR 8502, Universit\'{e}s Paris-Sud et Paris-Saclay, 91405 Orsay Cedex, France}

\date{\today}

\maketitle
\textbf{Magnetic skyrmions are textures behaving as quasiparticles which are topologically different from other states. Their discovery in systems with broken inversion symmetry sparked the search for materials containing such magnetic phase at room temperature. Their topological properties combined with the chirality-related spin-orbit torques make them interesting objects to control the magnetization at nanoscale. Here we show that a pair of coupled skyrmions with the same topological charge and opposite chiralities can be stabilized in a symmetric magnetic bilayer system by combining Dzyaloshinskii-Moriya interaction (DMI) and dipolar coupling effects. This effect opens a new path for skyrmion stabilization with much lower DMI. We then demonstrate in a single device with two different electrodes that such skyrmions can be efficiently and independently written and shifted by electric current at large velocities. The skyrmionic nature of the observed quasiparticles is further confirmed by using the gyrotropic force as a topological filter. These results set the ground for emerging spintronic technologies where issues concerning skyrmion stability, nucleation, and propagation are paramount.}

Over the past two years, a concerted effort has been made worldwide to study how magnetic skyrmions, a novel chiral phase evidenced in materials with broken inversion symmetry\cite{muhlbauer2009skyrmion,bogdanov1989thermodynamically,rossler2006spontaneous,heinze2011}, can appear and be displaced in ultrathin ferromagnetic films and nanotracks\cite{Fert2013,sampaio2013nucleation}. From an experimental viewpoint, there remain three important challenges that are determinant for whether skyrmions will be useful in future data storage technologies. First, the ability to tailor the chirality and energy of domain walls (DWs), such that isolated skyrmions remain sufficiently stable at room temperature against the ferromagnetic ground state. Second, it is important to drive the skyrmions efficiently with spin-transport torques, such as the spin Hall effect (SHE), as the reproducible displacement of skyrmions is crucial to information transfer. Third, it is essential to be able to nucleate skyrmions readily, since the ability to write (new) information is primordial to any storage application. Experimental demonstrations of certain aspects of these three points have been reported\cite{heinze2011,romming2013writing,Jiang2015,moreau-luchaire2016,woo2016observation,boulle2016}, however, independent writing and shifting in a single device remains a challenge.

Recently, significant developments in spintronics have being directly connected with broken inversion symmetry. On one hand, this allows spin-orbit torques to improve the efficiency of current induced magnetization manipulation\cite{miron2011,liu2012,emori2013current,ryu2013chiral}. On the other hand, such a situation allows to tailor DW chirality and lower the DW energy through the Dzyaloshinskii-Moriya interaction (DMI)\cite{dzyalo1965,heide2008dzyaloshinskii,hrabec2014measuring}, a requirement that ultimately permits the stabilization of skyrmions and their use in spintronics\cite{bogdanov1989thermodynamically,rossler2006spontaneous,Fert2013}. In general, the search for skyrmion host media requires a large DMI, which restricts the choice of materials. Here, we show that a globally symmetrical situation in magnetic bilayers meets all the requirements to host skyrmions, without the need for a very large DMI, since the control of DW chirality and energy is assisted by dipolar coupling. It results in two superimposed skyrmions, strongly coupled through their dipolar stray field, which behave as a single particle called skyrmion hereafter for simplicity. This method, compatible with spin-orbit torque-induced dynamics, offers a larger flexibility for the choice of materials. We show for the first time a simple and elegant technological way to independently write and shift such skyrmions at large velocities in a single functional device by means of electric current.

\begin{figure*}
  \includegraphics[width=15cm]{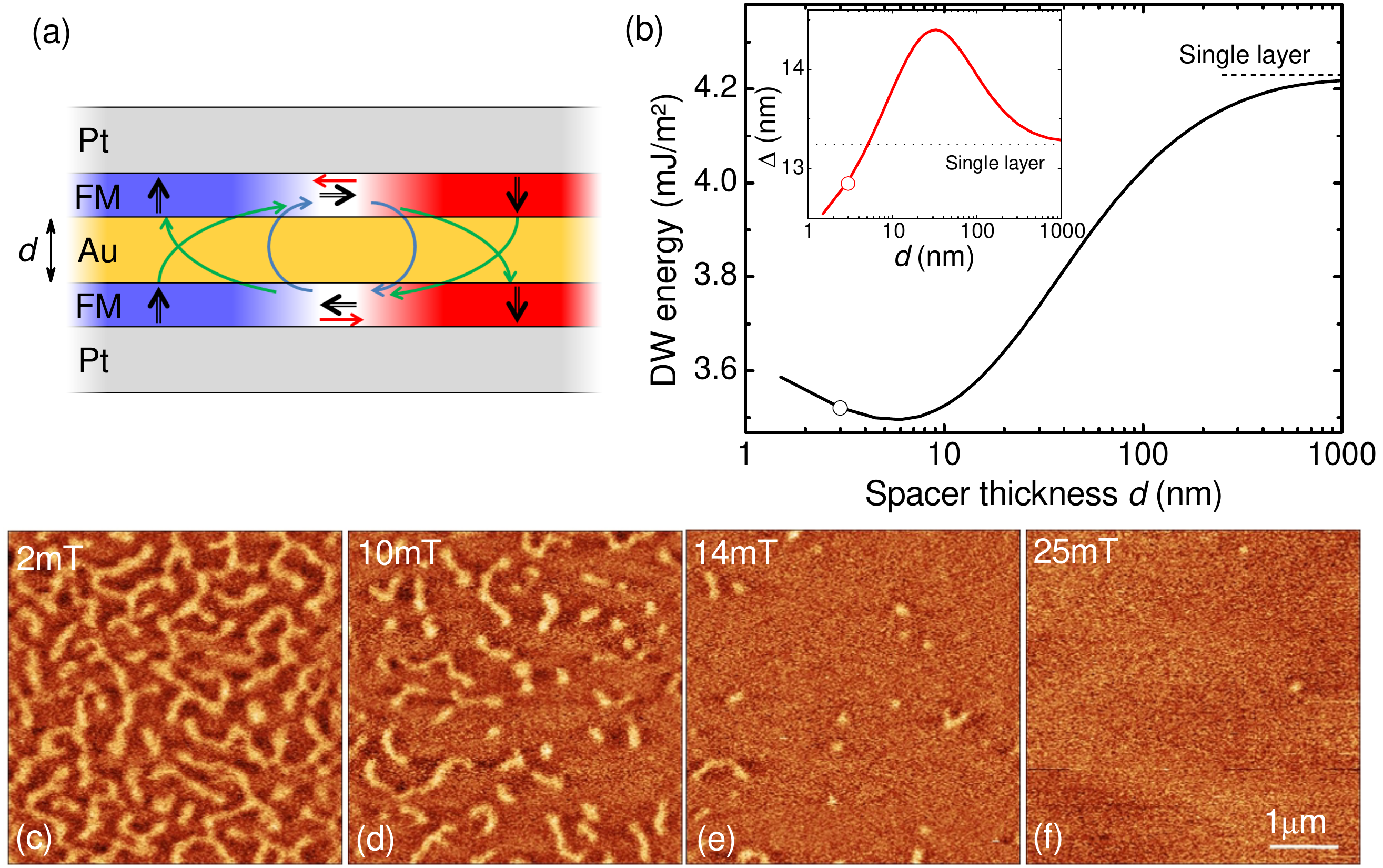}
  \caption{\textbf{Skyrmion stabilization.}
  (a) Sketch of the Pt/FM/Au/FM/Pt stack. The black arrows indicate magnetization orientation inside the two layers containing a DW. The coloured arrows correspond to DW internal (red),DW-DW (blue) and domain-DW (green) magnetostatic interactions respectively. (b) Calculated DW energy and width $\Delta$ as a function of the Au spacer thickness $d$. The circles correspond to the experimental case $d=3$~nm. (c)-(f) Field-dependent MFM images revealing the process of skyrmion formation. The bright and dark contrast corresponds to repulsive and attractive force respectively. The sample is in a demagnetized state at low field (c). The remagnetization with increasing field leads to condensation of skyrmions and a decrease of their density.
  \label{Fig_Energies}}
\end{figure*}

Stabilization of isolated skyrmions requires a fine control of the DW energy\cite{rohart2013PRB} between two limiting cases: a large positive energy causes skyrmions to collapse while a large negative wall energy destabilizes the collinear order and requires high magnetic fields to access the isolated skyrmions\cite{romming2013writing}. While this is generally achieved using DMI, we show that dipolar coupling in bilayers with perpendicular magnetization can be successfully used too. As illustrated in Fig.~\ref{Fig_Energies}(a) on a magnetic bilayer, the stray field arising from the domains couples to the DW magnetization, in a flux-closure configuration, and promotes N\'{e}el walls with opposite chirality in both layers and thus lowers the DW energy\cite{bellec2010domain}. The strength of this effect can be easily tuned by adjusting the spacer thickness, offering an additional means of control. The symmetric superimposition of two magnetic layers, each of them in a non-symmetric stacking, allows to satisfy both dipolar coupling and DMI. As the sense of the magnetization rotation induced by the stray field cannot be changed (left and right-handed in the bottom and top layer, respectively), the large spin-orbit layers generating the interfacial DMI should be placed as spacer if the induced DMI parameter $D$ is positive, or as outer layers if it is negative. Another advantage of bilayers is an increase of the dipolar interaction between the skyrmion core and the ferromagnetic surrounding as compared to a single layer, which further stabilizes skyrmions. As demonstrated below, this structure is also compatible with SHE-induced motion of skyrmions.

We use a stack of Pt(5~nm)$\backslash$FM$\backslash$Au($d$)$\backslash$FM$\backslash$Pt(5~nm) where FM=Ni$\backslash$Co$\backslash$Ni. The chosen FM gives the opportunity to change the surrounding metals without significantly changing the anisotropy which arises predominantly from the Ni$\backslash$Co interfaces\cite{bloemen1992magnetic}. Moreover, the use of two different ferromagnets gives more freedom to tune the stray field and the anisotropy by adjusting the Co and Ni thicknesses. Magnetometry and Brillouin light scattering spectroscopy on single magnetic layers with both stacking orders has shown that both layers are similar, which guaranties  film symmetry, only the sign of the DMI constant being opposite ($D =-0.21\pm0.01$~mJ/m$^2$ for Pt/FM/Au and $D=+0.24\pm0.01$~mJ/m$^2$ for Au/FM/Pt stacks respectively), as expected (see Supplementary Material). The DMI induced at the Pt/Ni interface thus yields a left-handed (counterclockwise) chirality\cite{chen2013tailoring} as sketched in Fig.~\ref{Fig_Energies}(a), which justifies the use of Pt layers as bottom and top layers, in order to satisfy the dipolar interaction. The thickness of Pt has been chosen with respect to the reported spin-diffusion length to maximize the SHE\cite{nguyen2016spin} and Au remains a neutral layer due to negligible SHE \cite{ryu2014chiral} and small DMI\cite{yang2015anatomy}. While exchange coupling has been already explored theoretically\cite{nandy2016interlayer} and experimentally\cite{chen2015room} for stabilizing skyrmions, here we focus on dipolar coupling. The thickness of Au is therefore chosen to avoid interlayer exchange coupling\cite{grolier1993unambiguous} but also to minimize the electric current shunting.

\section*{Skyrmion stabilization using dipolar couplings}

To understand the mechanism of the isolated skyrmion stabilization, one has to disentangle and quantify several energies involved in this process. The specific DW energy (hereafter called simply energy) $\sigma_0$, including the exchange and anisotropy energies\cite{hubertandschaefer}, is lowered by $\pi D$ upon introduction of DMI\cite{Thiaville_DMI,heide2008dzyaloshinskii}. The DMI favors a N\'{e}el wall structure, which also gives rise to magnetostatic charges on either side of the wall and creates a field opposed to the wall magnetization [red arrows in Fig.~\ref{Fig_Energies}(a)]\cite{Thiaville_DMI}. This is expressed by the energy increase $\delta_\mathrm{N}$. Alongside these usual energy terms, in the case of the bilayer system one has to take into account additional energies. The DW-DW magnetostatic interaction between two DWs with opposite chiralities leads to another energy term $\delta_\mathrm{DW-DW}$ which is illustrated by blue arrows in Fig.~\ref{Fig_Energies}(a). For spacer thicknesses $d$ smaller than the DW width $\Delta$, the magnetostatic charges created by each wall are so close that $\lvert\delta_\mathrm{N}\rvert\approx\lvert\delta_{\mathrm{DW-DW}}\rvert$ so that these two contributions almost compensate. On the contrary, the stray field arising from the domains, depicted by the green arrows in Fig.~\ref{Fig_Energies}(a) gives another significant energy decrease, $\delta_\mathrm{D-DW}$, which scales\cite{tetienne2015nature} approximately as $1/d$ and reinforces the chiral nature of each wall\cite{bellec2010domain}. The DW energy within such system then reads:
\begin{linenomath*}
 \begin{equation}\label{equation_energies}
\sigma\cong\sigma_0-\pi D  - \delta_\mathrm{D-DW}.
\end{equation}
\end{linenomath*}
The quantitative analysis of the energies has been performed by micromagnetic calculations. Fig.~\ref{Fig_Energies}(b) shows the DW energy dependence on the spacer thickness. The DW energy increases with increasing $d$ as expected from the argument of $1/d$ dependence. However, when the spacer thickness $d$ is comparable with the DW width shown in the inset of Fig.~\ref{Fig_Energies}(d), the DW energy decreases with $d$. The DW width strongly responds to the spacer thickness to accommodate the important energy changes. One can define the optimum spacer thickness  $d_{\mathrm{opt}}$ where the effect of dipolar energy contribution is maximized. With our parameters, this analysis yields $d_{\mathrm{opt}}$ close to 3~nm, and the ratio between the DMI and magnetostatic energies is $\delta_\mathrm{D-DW}/\pi D=1.1$, i.e. half of the energy minimization is due to the magnetostatic energy. This approach based on isolated DWs can be also extended to $360^{\circ}$ DWs and skyrmions where the stray field is slightly lower, showing that the dipolar mechanism is efficient to stabilize skyrmions down to at least 20~nm (see Supplementary Material).

While the DW energy remains positive here, the dipolar coupling between the skyrmion core and the ferromagnetic surrounding can efficiently lower the skyrmion energy cost. However, in ultrathin films, dipolar coupling vanishes with the thickness\cite{kooy1960experimental} so that the film thickness must be larger than a characteristic length $l_\mathrm{c}=\sigma/\mu_0M_\mathrm{s}^2$ to spontaneously promote magnetic textures\cite{kooy1960experimental,boulle2016} (with $M_\mathrm{s}$ the spontaneous magnetization). In our case, $l_\mathrm{c}=3.9$~nm so that a single layer film (1.5~nm thick) can hardly be demagnetized and shows a full remanence square hysteresis loop. On the contrary, bilayer films have a total thickness larger than $l_\mathrm{c}$ and therefore spontaneously demagnetize in a multiple domain state, as shown by a zero remanence hysteresis loop (see Supplementary Material).

The magnetic texture imaged by magnetic force microscopy (MFM, see Methods) confirms that the bilayers are at remanence in a worm-like demagnetized state. Figs.~\ref{Fig_Energies}(c)-(f) show a sequence of MFM images at different fields and illustrate how the worm-like structure can be unwound into the isolated skyrmion phase with moderate fields. As soon as an out-of-plane magnetic field is applied the domains start to contract into skyrmions whose size and density decrease with the magnetic field. Close to the saturation field only a few isolated skyrmions remain. They are the ones needed for free skyrmion dynamics studies and applications. We note that the MFM does not provide any insight into the topology of the created textures, a point addressed later. In the following, we focus on skyrmions in nanostructures. In those, size effects lower the magnetic field needed to saturate the sample so that applying only 6~mT is sufficient to study isolated skyrmions of $160\pm40$~nm diameter, measured within the accuracy of the MFM (see Fig.~\ref{Fig_racetrack}).

\section*{Skyrmion nucleation}
\begin{figure*}[!ht]
  \includegraphics[width=18cm]{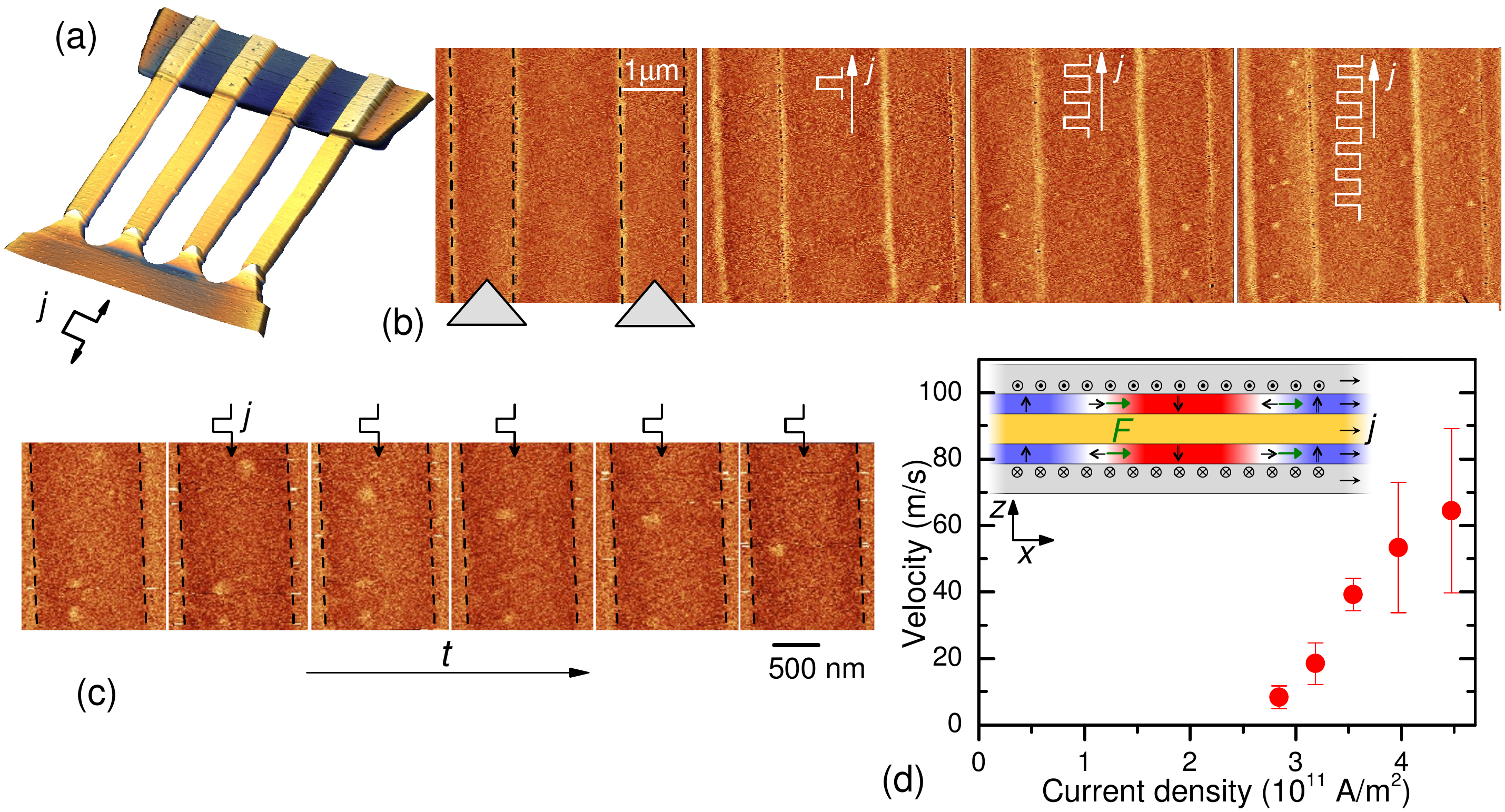}
  \caption{\textbf{Skyrmion generation and dynamics.} (\textbf{a}) AFM image of the asymmetric device with the skyrmion injection tips. (\textbf{b}) Skyrmion writing by at $B_z=-6$~mT by 7~ns long, $j=2.8\times10^{11}$~A/m$^2$ pulses starting from a fully saturated state, with one injected skyrmion after application of first pulse, and several skyrmion after a train of pulses. The dashed lines and triangles correspond to the wire edges and electric contacts respectively. \textbf{(c)} Series of images showing skyrmion shift along the track between 3~ns, $j=3.9\times10^{11}$~A/m$^2$ electric pulses. \textbf{(d)} Measured velocity of skyrmions as a function of current densities at $B_z\simeq-6$~mT. The inset shows a sketch of the skyrmion cross section with the spin accumulation due to the electrical current $\vec{j}$. The resulting force $\vec{F}$ defined by Equation~(\ref{equation_force}) causes both skyrmions to move in the same direction (against the electron flow).\label{Fig_racetrack}}
\end{figure*}

Several methods have been proposed to nucleate skyrmions\cite{iwasaki2013current,sampaio2013nucleation,Jiang2015,romming2013writing,everschor2016skyrmion}. Here we demonstrate another method which is characterized by its simplicity and integrability. To study the skyrmion dynamics in confined structures we have fabricated the device shown in Fig.~\ref{Fig_racetrack}(a) containing four parallel, 1~$\mathrm{\mu}$m wide wires. The electrical contacts are fabricated in a non-symmetric fashion where one side of the wires is connected by sharp tips and the other by a wide electrode. At the tip current lines divergence\cite{heinonen2016generation}, heating, and spin-accumulation\cite{gorchon2014stochastic} are largest, disturbing the magnetic configuration. As a result, skyrmions spontaneously appear there and are injected into the track. Fig.~\ref{Fig_racetrack}(b) shows a systematic skyrmion nucleation in two adjacent wires in the vicinity of the contact. Skyrmions are injected into fully saturated wires (-6~mT applied field) by a series of $7$~ns long pulses. Above a threshold of $j_\mathrm{c}\simeq2.6\times10^{11}$~A/m$^2$ the skyrmions are injected at the tip, as shown after application of the first pulse, and carried away into the tracks. When the polarity of the electric current is reversed this injection mechanism does not work, as the nucleated skyrmions are pushed by the current toward the tip. Nucleation at the wide electrode has never been observed. Therefore due to this geometrical asymmetry one polarity of the electric current serves as a skyrmion generator while the other simply shifts the existing skyrmions.

\section*{Skyrmion dynamics}

After filling the tracks with skyrmions we reverse the polarity of the current to avoid interaction with newly injected skyrmions, and so switch to the skyrmion shifting mode. To measure the velocities as a function of current density, we measure the skyrmion displacement after application of each current pulse, of duration ranging from 3 to 10~ns. Fig.~\ref{Fig_racetrack}(c) shows a sequence of images demonstrating skyrmion displacement by 3~ns long pulses with a current density $j=3.9\times10^{11}$~A/m$^2$. The resulting measured skyrmion velocity as a function of current densities at $B_z=-6$~mT presented in Fig.~\ref{Fig_racetrack}(d) reveals skyrmion velocities up to 60~m/s. From one pulse to another, even at the largest current densities, successive displacement lengths are not equal, which underlines the role of defects and that skyrmions move by hopping within the potential landscape\cite{kim2009interdimensional}. They advance in the track until they get pinned, or reach a strongly pinned skyrmion which prevents further propagation by skyrmion-skyrmion repulsion.

The skyrmions move in the direction opposite to the electrons which suggests that it is the spin Hall effect that governs their dynamics\cite{emori2013current,sampaio2013nucleation}. When an electric current passes through the Pt layers, a spin accumulation with opposite polarities is generated at each interface, as sketched in Fig.~\ref{Fig_racetrack}(d). The force acting on a skyrmion can be expressed as\cite{sampaio2013nucleation}
\begin{linenomath*}
\begin{equation}\label{equation_force}
\vec{F}_{\mathrm{SH}}=\pm\frac{\hbar }{2 e} \pi j \theta_{\mathrm{SH}} b \vec{e}_z \times \vec{e}_\mathrm{p}
\end{equation}
\end{linenomath*}
where $j$ is the current density, $\theta_{\mathrm{SH}}$ is the SHE angle and $b$ is a skyrmion characteristic length (half its perimeter when the skyrmion radius $R$ is much larger than the DW width $\Delta$). The SHE-induced spin accumulation is along the vector $\vec{e}_\mathrm{p}=\vec{n} \times \vec{j}$, with $\vec{n}$ being the outer normal to the SHE layer at the interface considered, and its sign is given by that of the SHE angle $\theta_\mathrm{SH}$ (positive for Pt\cite{nguyen2016spin}). With $\vec{e}_z$ the vertical direction in the laboratory frame (for example, from substrate towards film), used to define the chirality of the skyrmions, the sign of the force is set by the chirality, as specified by the $\pm$ symbol in Eq.~\ref{equation_force} where ``$+$'' stands for right-hand chirality. Since the spin accumulation and the chirality are both opposite at each interface, the forces acting on the skyrmions depicted in Fig.~\ref{Fig_racetrack}(d) point in the same direction in both magnetic layers. The skyrmions in both layers are therefore pushed in the same direction, along the electrical current here. The obtained velocities quantitatively agree with the model for free skyrmion dynamics in a disorder-free medium (see Supplementary Material).

\begin{figure*}
  \includegraphics[width=16cm]{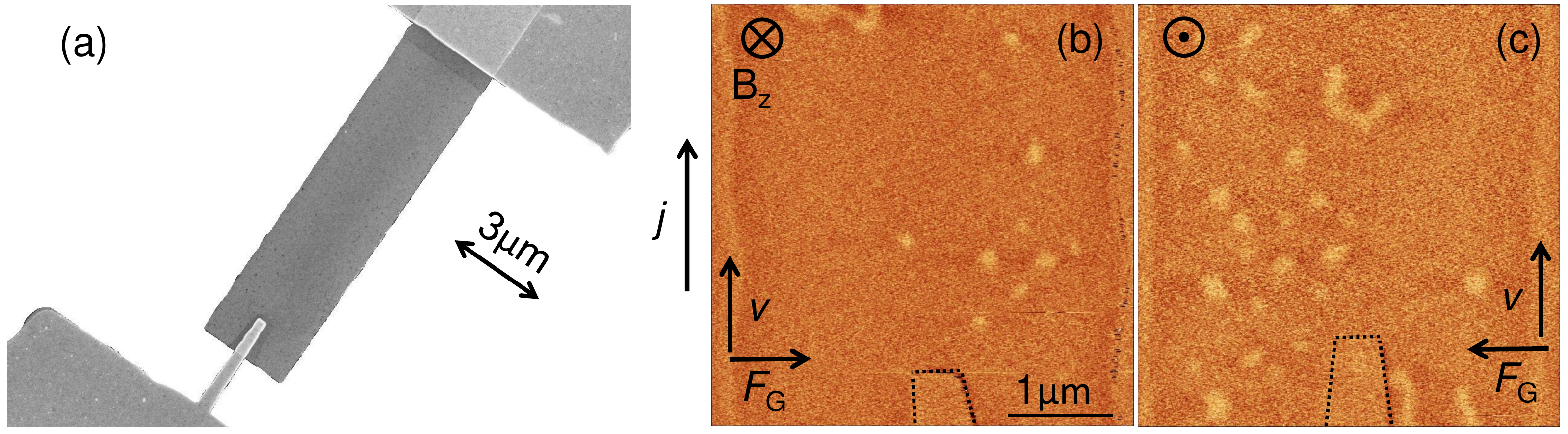}
  \caption{\textbf{The gyrotropic force.} (\textbf{a}) SEM micrograph of the sample geometry used for skyrmion generation demonstration. The 3~$\mathrm{\mu}$m ferromagnetic stripe is connected to Ti$\backslash$Au pads with an injection finger 300~nm wide. (\textbf{b}) Resulting state after application of a series of 8~ns long pulses in an external field of $B_z\simeq-6$~mT where $j=3\times10^{11}$~A/m$^2$. (\textbf{c}) The same experiment with $B_z\simeq+6$~mT shows skyrmion accumulation on the left side of the device. The dashed line indicates position of the Ti/Au finger. \label{Fig_Magnus}}
\end{figure*}
\section*{Skew deflection: insight into topology}

One way to confirm that we deal with topological textures is to prove that they experience a gyrotropic force\cite{thiele1973steady}, expressed as
\begin{linenomath*}
\begin{equation}\label{equation_gyro}
\vec{F}_\mathrm{G}=\vec{G}\times \vec{v} = \left(0,0, -\frac{\mu_0 M_\mathrm{s} t}{\gamma_0}\Omega\right) \times (v_x,v_y,0)
\end{equation}
\end{linenomath*}
where $\vec{G}$ is the gyrotropic vector, $\Omega=4\pi S p$ with $S$ being the winding number and $p$ the core polarity. $\Omega$ thus intimately binds the topology of the quasiparticle to its motion, providing a way to reveal its topological state. The skyrmion dynamics can be described by the massless Thiele equation\cite{thiele1973steady,thiele1974applications}
\begin{linenomath*}
\begin{equation}\label{equation_Thiele}
\vec{G} \times \vec{v} - \alpha \overset{\longleftrightarrow}{\mathbf{\mathcal{D}}} \cdot \vec{v} + \vec{F}_{\mathrm{SH}} = 0
\end{equation}
\end{linenomath*}
where $\vec{F}_{\mathrm{SH}}$ is the force expressed by Equation~(\ref{equation_force}), $\alpha$ the Gilbert damping and $\overset{\longleftrightarrow}{\mathbf{\mathcal{D}}}$  the dissipative tensor (see Supplementary Material for its calculation). The gyrotropic force therefore deviates the skyrmions from the direction of $\vec{F}_{\mathrm{SH}}$ in a similar manner as electrons moving in a magnetic field\cite{tomasello2013strategy,jiang2016direct,litzius2016skyrmion}.

To study this behavior we have lifted the geometrical constrain and patterned the ferromagnetic stack into a 3~$\mathrm{\mu}$m wide strip where the skyrmions can move freely in the lateral direction. The electrical contact is designed asymmetrically by a 300~nm wide non-magnetic finger and a straight electrode as shown in Fig.~\ref{Fig_Magnus}(a). We have verified that in this geometry the skyrmions can be only generated by one polarity of the current similarly to the case shown in Fig.~\ref{Fig_racetrack}(b). The ferromagnet was first saturated and then a series of 8~ns long, $j=3\times10^{11}$~A/m$^2$ current pulses at $B_z=-6$~mT has been applied in order to inject the skyrmions. Fig.~\ref{Fig_Magnus}(b) demonstrates a resulting magnetic configuration after application of a series of pulses showing several skyrmions. The skyrmions are ejected at the tip and are carried away by the current with a certain deflection towards right. Note that this deflection is also visible in Fig.~\ref{Fig_racetrack}(c) (to the left as the current is opposite). In order to prove the behavior predicted by Equation~(\ref{equation_gyro}) we have generated skyrmions with opposite core magnetization by applying $B_z=+6$~mT, i.e. changing the sign of $\Omega$. Fig.~\ref{Fig_Magnus}(c) shows a resulting skyrmion configuration after application of a train of pulses of the same polarity as in Fig.~\ref{Fig_Magnus}(b). In this case, the skyrmions accumulate on the left side of the sample. We emphasize that the direction of the magnetic field shown in Fig.~\ref{Fig_Magnus}(b) and (c) is absolute and the skyrmion accumulation indeed appears on the side predicted by Equation~(\ref{equation_gyro}). The effect of Oersted field can be excluded as it would give the opposite deflection direction (see Supplementary Material). This undoubtedly confirms that our textures have $S>0$ topology compatible with skyrmions. A quantitative determination of the winding number is impossible as the track width and the skyrmion trajectory between the imperfections limit the deflection\cite{jiang2016direct,reichhardt2015collective,Kim_CurrentDriven}. Note that $S=1$ state would imply a deflection angle of $\approx73^{\circ}$ (see Supplementary Material).

To conclude, we have engineered a magnetic bilayer system which efficiently employs all the available energies (DMI and dipolar couplings) to stabilize a pair of skyrmions. The two skyrmions have the same topological charge while having opposite chiralities and are strongly coupled through their dipolar stray field. The opposite chirality in combination with a reversed spin accumulation results in a system suitable for the current-induced dynamics. We have developed functional devices with asymmetric electrodes designed for systematic current-induced skyrmion generation and motion. The skyrmion nature of the quasiparticles was demonstrated  by topological filtering employing the gyrotropic force. The guideline used here widens the possibilities of skyrmion stabilization and manipulation and, hence, their applications.

\nolinenumbers
\section*{Methods}
\begin{footnotesize}
\subsection*{Sample preparation}
Samples were grown in an ultra-high vacuum evaporator with base pressure of $10^{-10}$~mBar. The multilayers of Pt$\backslash$FM$\backslash$Au$\backslash$FM$\backslash$Pt with FM=Ni(4~{\AA})$\backslash$Co(7~{\AA})$\backslash$Ni(4~{\AA}) were deposited on a high-resistive silicon with native oxide layer in order to minimize the Joule heating effect during electric current pulse application\cite{torrejon2012unidirectional}. The symmetry of the stack has been verified by magnetometry measurements on Pt/FM/Au and Au/FM/Pt stacks using SQUID ($M_\mathrm{s} = 0.85$~MA/m, anisotropy field of about 150~mT), and Brillouin light scattering spectroscopy (BLS) to determine DMI\cite{di2015direct} (see Supplementary Material). The films were patterned by e-beam etching using an aluminium hard mask which was consequently removed by chemical etching. The Ti$\backslash$Au contacts were made in the second step via lift-off technique.

\subsection*{Magnetization imaging and dynamics}
MFM was performed on a commercial Bruker Dimension 3000 microscope with a stage customized for high frequency transport measurements with a typical pulse rise/fall time $<1$~ns\cite{chauleau2010magnetic}. The MFM tips are home made with a non-magnetic capping layer in order to increase the distance between the magnetic tip coating and the surface to minimize the magnetic perturbations during the AFM scan. Due to the fact that the tips are magnetically extremely soft and so follow the applied field, the skyrmions always appear as a bright (i.e. repulsive) contrast due to the antiparallel magnetic configuration between the tip and the skyrmion. To avoid any heating and related thermal drifts, we have used permanent magnet which implies an error of 20\% on the given values of magnetic field. Current densities are defined as the average across the entire sample thickness (magnetic and non-magnetic layers).

\subsection*{Micromagnetic calculations}
Micromagnetic modelling was carried out using the OOMMF code where the cell size used was 0.5~nm $\times$ 1.5~nm $\times$ 1.5~nm. We used the measured material parameters $K_\mathrm{u}=0.5$~MJ/m$^{3}$, $M_{\mathrm{s}}=0.85$~MA/m, and used $A=12$~pJ/m as an average of the bulk exchange constants for Co and Ni.
\end{footnotesize}


\bibliographystyle{naturemag}

\vskip 1cm
\begin{footnotesize}
\section*{Acknowledgements}
We thank to Joo-Von Kim and Jacques Miltat for critical reading of the manuscript. This work has been supported by the Agence Nationale de la Recherche (France) under Contract No. ANR-14-CE26-0012 (Ultrasky), ANR-09-NANO-002 (Hyfont), the RTRA Triangle de la Physique (Multivap), the European Research Council (ERC-StG-2014, Imagine) and by the Government of the Russian Federation, (Grant 074-U01).

\section*{Author Contributions}
S.R., A.H. and A.T. conceived the study. A.H. and S.R. deposited and characterized the multilayer samples. A.H. and R.W. patterned the samples. A. H. performed the MFM measurements. S.R., J.S. and A.T. performed the micromagnetic simulations. M.B. participated in performing the BLS measurements and analyzed all the BLS data. All authors discussed the data and reviewed the manuscript.

\section*{Additional Information}
\textbf{Competing financial interests:} The authors declare no competing financial interests.

\textbf{Reprints and permission} information is available online at http://npg.nature.com/reprintsandpermissions.
\end{footnotesize}

\widetext
\clearpage
\setcounter{equation}{0}
\setcounter{figure}{0}
\setcounter{table}{0}
\renewcommand{\theequation}{S\arabic{equation}}
\renewcommand{\thefigure}{S\arabic{figure}}

\section{Sample characterization}

The sample symmetry is a key point in this work. In order to check our ability to grow the magnetic films with similar parameters for both Pt$\backslash$FM$\backslash$Au and Au$\backslash$FM$\backslash$Pt stack order, we have characterized samples containing only one magnetic layer. All films were grown on a Ta(3~nm)$\backslash$Pt(5~nm) buffer layer deposited on Si(001) with a native oxide layer. For the Pt$\backslash$FM$\backslash$Au sample, the FM [Ni(4~{\AA})$\backslash$Co(7~{\AA})$\backslash$Ni(4~{\AA})] layer has been grown directly on the buffer layer, while for the Au$\backslash$FM$\backslash$Pt, a 5 nm thick Au layer was grown before the FM. The cover layer (Au and Pt respectively) was 5~nm thick. These samples were compared to the magnetic bilayer sample Pt$\backslash$FM$\backslash$Au($d$)$\backslash$FM$\backslash$Pt with $d=3$~nm and $d=5$~nm.

Using SQUID, the saturation magnetization $M_\mathrm{s}=0.85\times10^6$~A/m has been determined on all single and bilayer samples, and is close to the value ($0.9\times10^6$~A/m) expected from the bulk magnetization of Co ($1.37\times10^6$~A/m) and Ni ($0.49\times10^6$~A/m).

\begin{figure}[h!]
\includegraphics[width=14cm]{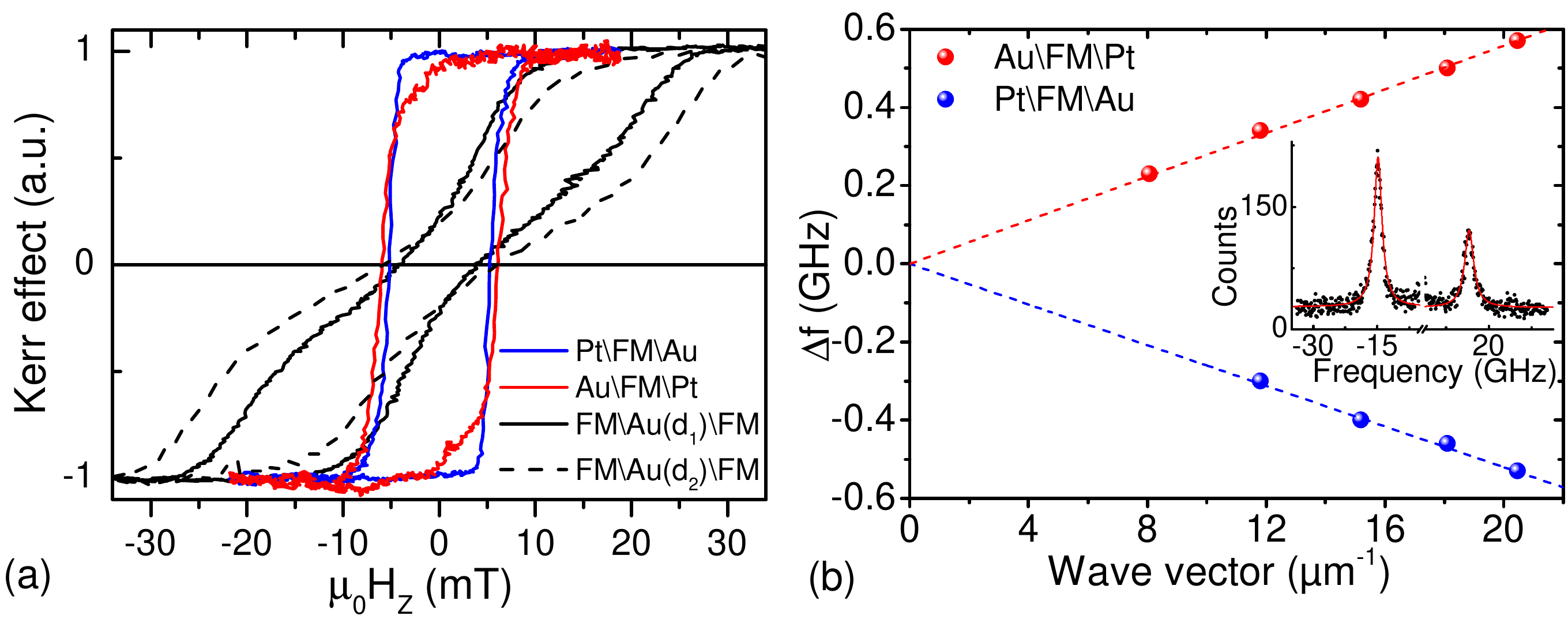}
\caption{\textbf{Stack characterization.} (a) Hysteresis loops of the single magnetic layer of Pt/FM/Au and Au/FM/Pt where FM$=$Ni(4~{\AA})$\backslash$Co(7~{\AA})$\backslash$Ni(4~{\AA}), and of corresponding magnetic bilayer with different thickness of Au spacer $d_1=5$~nm and $d_2=3$~nm. (b) Frequency shift $\Delta f = f_\mathrm{S} - f_\mathrm{AS}$ obtained in Pt/FM/Au (blue) and Au/FM/Pt (red) stacks. The inset shows an example of BLS spectra measured at $20.45$~$\mathrm{\mu}$m$^{-1}$ in Pt/FM/Au, under a 0.6~T in-plane field, with red lines corresponding to Lorentzian fits. \label{Fig_Char}}
\end{figure}

Fig.~\ref{Fig_Char}(a) shows hysteresis loops determined using magneto-optical polar Kerr effect, as a function of an out-of-plane magnetic field for both single layer samples and two bilayer samples which differ by the spacer thickness $d$. While both single layers show similar square hysteresis, i.e. full remanence and similar coercive field, the bilayer samples have a lower remanence indicating spontaneous tendency to demagnetization. Note that the thicker the spacer, the smaller the saturation field, which indicates a decreased coupling.

The DMI has been quantified by the Brillouin light scattering (BLS) method \cite{di2015direct,belmeguenai2015interfacial}, in Damon-Eshbach geometry (in-plane magnetization, spin wave propagation vector perpendicular to the magnetization). Here two oppositely propagating spin waves defined by wave vector $k_x$ are sensitive to the DMI parameter $D$ causing a frequency shift $\Delta f$ between Stokes (spin-wave creation) and anti-Stokes (spin-wave annihilation) modes. An example of Stokes and anti-Stokes peaks for an in-plane magnetic saturating field $\mu_0H_y=0.6$~T is displayed in Fig.~\ref{Fig_Char}(b). To access the DMI in the individual layers, we have measured the $\Delta f(k_x)$ dependence on single Pt/FM/Au and Au/FM/Pt layers. The difference between Stokes and anti-Stokes frequencies $\Delta f = f_\mathrm{S} - f_\mathrm{AS}=2\gamma k_x D/ \pi M_\mathrm{s}$, shown in Fig.~\ref{Fig_Char}(b) allows extraction of the DMI parameter from the linear fits. We find effective values of $D =-0.21\pm0.01$~mJ/m$^2$ for Pt/FM/Au and $D=+0.24\pm0.01$~mJ/m$^2$ for Au/FM/Pt stacks, respectively.

Additionally, BLS spectra provide an estimation of the magnetic anisotropy and the damping factor. The resonance frequency is $f_\mathrm{r}=\sqrt{f_xf_z}$, where the two principal frequencies read $f_x=(\gamma_0/2\pi)H_y$ and $f_z=(\gamma_0/2\pi)(H_y-H_\mathrm{K})$, where $H_\mathrm{K}$ is the effective anisotropy field. These two frequencies shift linearly in $k_x$ in the presence of DMI, but the shift is small so we perform  calculation at zero wave vector. From  $\mu_0H_y=0.6$~T and the gyromagnetic factor $g=2.17$ for Co, one estimates $f_x=18.2$~GHz, hence from the experimental $f_\mathrm{r}=15.75$~GHz, defined as the mean between Stokes and anti-Stokes frequencies, one gets $f_z=13.6$~GHz. The anisotropy field is then $\mu_0H_\mathrm{K}=152$~mT. Subtracting the shape anisotropy $K_d=\frac12\mu_0M_\mathrm{s}^2$, the sample anisotropy is 0.52~MJ/m$^3$. From the expression of the imaginary part of the magnetic susceptibility one gets, for small damping, that the full width at half-maximum (FWHM, in the frequency domain)  $\delta f$ of the BLS peak is $\delta f=\alpha(f_x+f_z)$. Assuming that no inhomogeneous broadening exists, the FWHM of the BLS peak thus provides an upper limit for the Gilbert damping $\alpha$. On the spectrum shown in Fig.~\ref{Fig_Char}(b) one measures $\delta f=2.5\pm0.3$~GHz. From this we find $\alpha\leq0.08\pm0.01$.

\begin{figure}[h!]
\includegraphics[width=7cm]{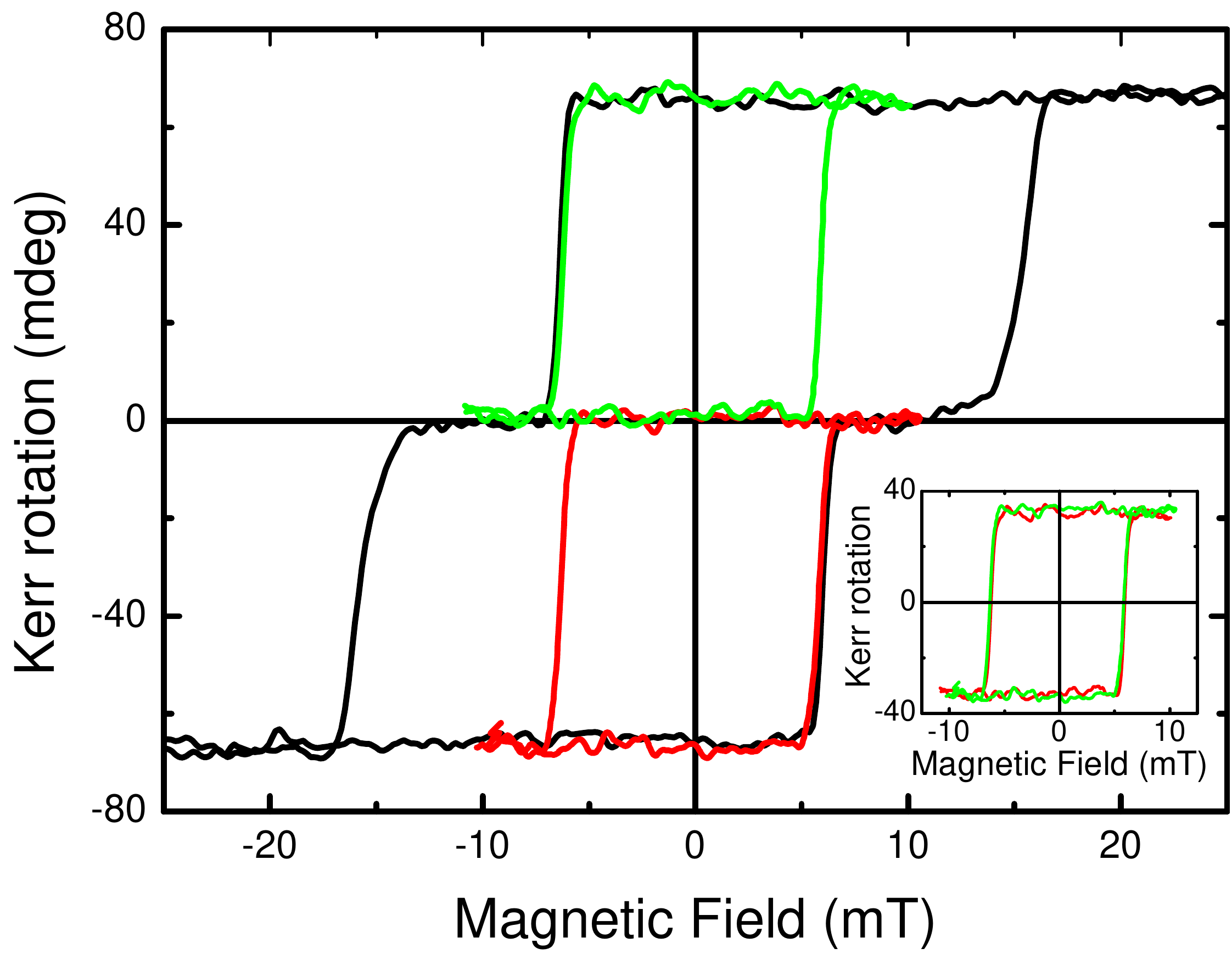}
\caption{\textbf{Interlayer coupling.} Full and minor hysteresis loops measured on Pt(5~nm)$\backslash$Co(5~{\AA})$\backslash$Ni(6~{\AA})$\backslash$Au($3$~nm)$\backslash$Ni(5~{\AA})$\backslash$Co(5~{\AA})$\backslash$Ni(5~{\AA})$\backslash$Pt(5~nm) by polar Kerr magnetometery. The inset shows superimposition of the two minor loops. \label{Fig_ElCoup}}
\end{figure}

In order to verify that the electronic coupling in our bilayer is eliminated by the choice of 3~nm Au spacer, we have grown an unbalanced sample with stronger perpendicular anisotropies in order to obtain two square hysteresis loops with different coercive fields: Pt(5~nm)$\backslash$Co(5~{\AA})$\backslash$Ni(6~{\AA})$\backslash$Au($3$~nm)$\backslash$Ni(5~{\AA})$\backslash$Co(5~{\AA})$\backslash$Ni(5~{\AA})$\backslash$Pt(5~nm). The major and minor hysteresis loops are shown in Fig.~\ref{Fig_ElCoup}. The two observed steps correspond to magnetization reversal in the first and second layer. The electronic coupling between both layers can be deduced from the bias field in the minor loops. Since no bias field is observed, as shown from the perfect superposition between minor loops obtained with a hard layer saturated up or down, the interlayer exchange coupling is negligible for a 3~nm spacer, in agreement with previous investigation on Co/Au/Co system\cite{grolier1993unambiguous}.

\begin{figure}[h!]
\includegraphics[width=5cm]{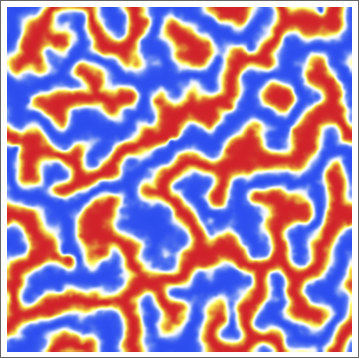}
\caption{\textbf{Micromagnetic simulations.} Calculated out-of-plane magnetization component of a demagnetized film with the measured parameters and $A=12$~pJ/m (see Methods for more details). Red and blue color corresponds to up and down magnetization component respectively. Size of the displayed box is $1.5~\mu\mathrm{m} \times 1.5~\mu\mathrm{m}$. Calculated by MuMax3 code\cite{Vansteenkiste2014} with periodic boundary conditions. \label{Fig_Sim}}
\end{figure}

We have simulated the relaxed state of the system using micromagnetic simulations with the experimentally measured parameters (see Methods for more detail). We have added a 5\% fluctuation in thickness affecting the $K_\mathrm{u}$, $M_\mathrm{s}$ and $D$ to account for pinning. The simulations, shown in Fig.~\ref{Fig_Sim}, reproduce well the magnetization pattern of the MFM data (Fig.~1(c) of the main text).

\section{Stabilization mechanism via dipolar coupling}

This paper proposes a new mechanism, based on dipolar couplings, for chiral DWs stabilization by lowering its energy, key towards skyrmion stabilization. While in the main text, and in particular in Fig.~1(b), the demonstration is performed on an isolated DW, we extend here the discussion to the more complex situations of $360^{\circ}$ DWs and skyrmions.

Dipolar couplings have been shown to lower the DW energy in the symmetric bilayer DWs, through a flux closure mechanism. This implies long range interaction of the stray field emitted by the domains with the DW magnetization. In a skyrmion situation, schematically, the strength of the interaction should reduce as the two walls on either side of the skyrmion core come close to each other, which lowers the stray field coming out from the skyrmion core. Here we show that this picture based on the isolated DWs remains unchanged down to 40~nm separated DWs.

\begin{figure}[h!]
\includegraphics[width=17cm]{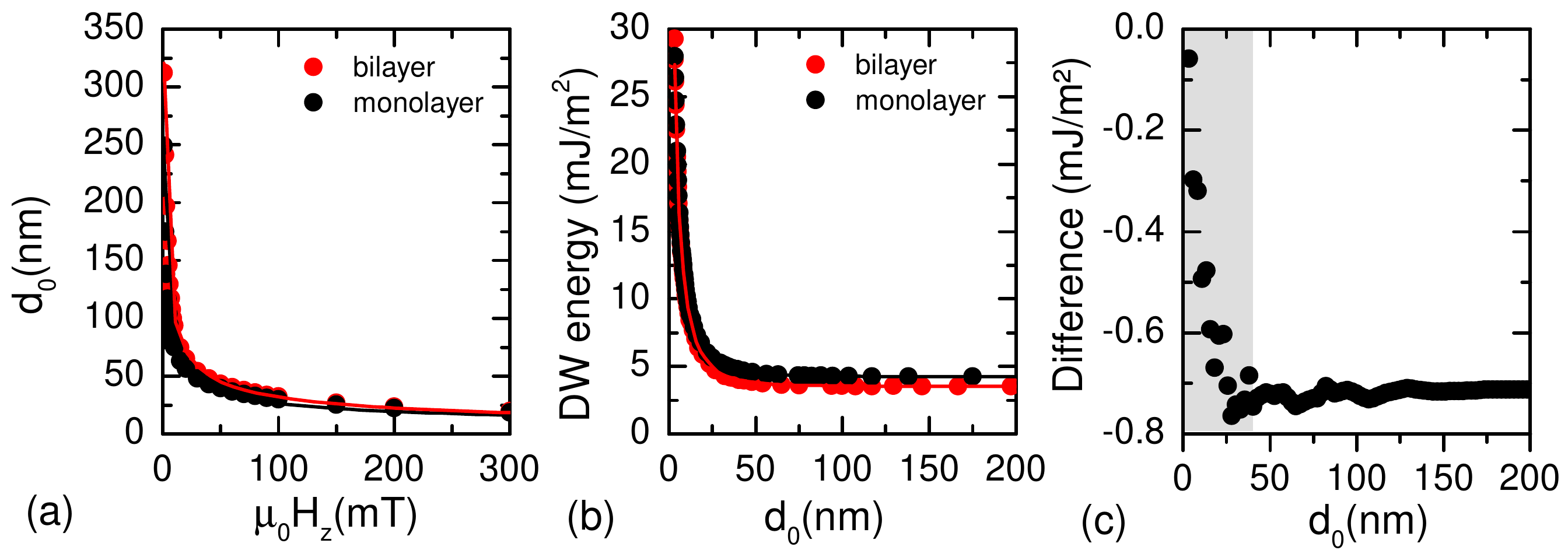}
\caption{\textbf{$\mathbf{360^{\circ}}$ DW stabilization} (a) $360^{\circ}$ DW separation $d_0$ as a function of the applied magnetic field for single and bilayer films. The dots correspond to the micromagnetic simulation and the lines to a $1/\sqrt{H_z}$ fit. (b) DW energy as a function of $d_0$. The full lines correspond to an interpolation used to calculate the energy difference. (c) DW energy difference between mono and bilayer films, which corresponds to the energy gain due to the flux closure mechanism. The low frequency variations are due to calculation and interpolation imprecisions and thus are meaningless. The decrease to zero below 40~nm (gray area) is due to the decay of the flux closure mechanism efficiency for small DW separations. \label{fig_360walls}}
\end{figure}

We first consider a $360^{\circ}$ DW. In such a situation, two chiral $180^{\circ}$ DWs are separated by a fixed distance $d_0$. The distance can be controlled by an external perpendicular magnetic field $H_z$, oriented against the magnetization of the domain situated between the two DWs, as $d_0\propto1/\sqrt{H_z}$ [\onlinecite{bauer2005deroughening}]. We have calculated the DW energy\cite{notedomainwallenergycalc} in single and bilayers (for a 3~nm thick spacer), as a function of $d_0$, as plotted in Fig.~\ref{fig_360walls}. For a large separation, we find the same DW energy as for isolated DWs. As $d_0$ decreases, we observe an increase of the DW energy below about 40~nm, for both mono and bilayer DWs. Such an increase cannot be attributed only to the decay of the dipolar coupling mechanism. Indeed, as the distance becomes comparable to the DW width, the DW profile adapts to the compression of the $360^{\circ}$ DW, so that all the micromagnetic energies are expected to increase. In single layers, the energy increase is only due to this effect so that this calculation serves as a reference. The DW energy gain due to the flux closure mechanism corresponds to the difference between the single and bilayer calculation. For large $d_0$, we find $\approx0.7$~mJ/m$^2$ energy gain in bilayers similar to the result in Fig.~1(b). This value is rather constant down to 40~nm separation then drops down for smaller separation. For a 10~nm distance, i.e. smaller than the DW width, the strength of flux closure effect is still 50\% of that at large separation.

\begin{figure}[h!]
\includegraphics[width=11cm]{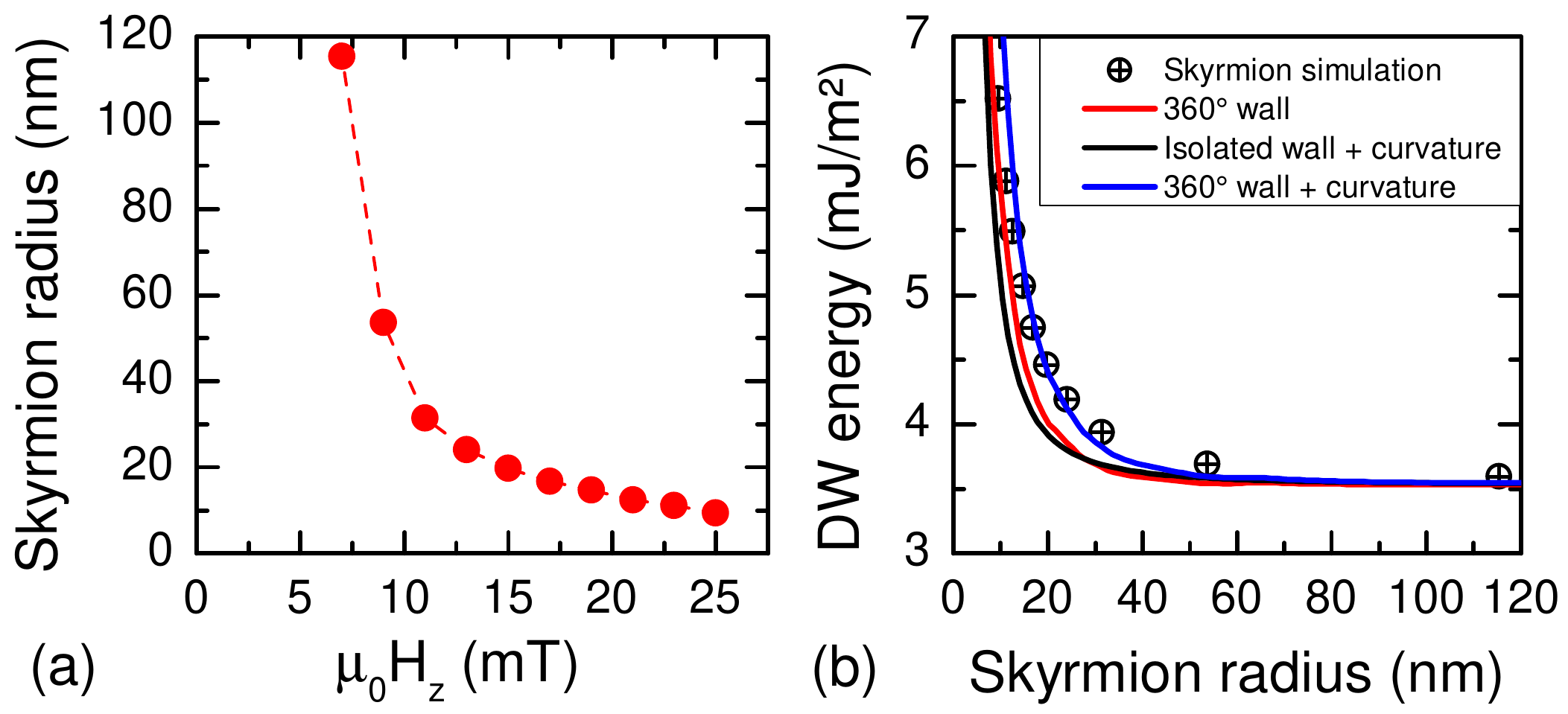}
\caption{\textbf{Skyrmion stabilization} (a) Skyrmion radius variation as a function of the external field strength. At 6~mT, a good agreement is found with the experiment. (b) Extracted DW energy as a function of the skyrmion radius. The dotted line corresponds to the DW curvature energy, as calculated from Ref.~\onlinecite{rohart2013PRB}, and the full line corresponds to the sum of the DW curvature energy and the DW energy as calculated form the $360^{\circ}$ DW study. \label{fig_skyrms}}
\end{figure}

Performing the same calculation for skyrmions is more complicated, as skyrmions could not be stabilized in monolayer films and therefore could not be used as a reference. However, we have calculated the compression of skyrmions under a perpendicular field $H_z$ oriented against the skyrmion core and extracted the DW energy\cite{notedomainwallenergycalc}. We find that, thanks to the external field, skyrmions can be compressed down to 20~nm diameter at 25~mT as shown in Fig.~\ref{fig_skyrms}(a). The extracted DW energy, displayed in Fig.~\ref{fig_skyrms}(b), is found to increase as the skyrmion is compressed. At large skyrmion diameter, the DW energy is in good agreement with the isolated DW energy. The increase at small diameter can be attributed to the interaction between diametrically opposite DWs, similarly to what has been discussed in $360^{\circ}$ DWs, as well as to the curvature of the DWs\cite{rohart2013PRB}. Using the curvature energy $\sigma_c\approx2A\Delta/r_\mathrm{s}^2$ from Ref.~\onlinecite{rohart2013PRB} and the DW energy extracted from the $360^{\circ}$ DW analysis, we reproduce quite well the DW energy. The difference can be attributed on one hand to the imprecision of the curvature energy formula (demonstrated for large skyrmions) and on the other hand to a further lowering of the flux closure mechanism presented in this study, as compared to $360^{\circ}$  DWs. However, it turns out that, whatever the decreased efficiency of the flux closure mechanism, such a mechanism still remains so that we are able to demonstrate, in the calculation, ultrasmall skyrmion stabilization. This opens perspectives for future experiments to stabilize smaller skyrmions. Tuning the micromagnetic parameters should allow ultrasmall skyrmions stabilization at smaller applied magnetic field than in this numerical study.

\section{Skyrmion deflection: Competition between gyrotropic and Oersted field effects}

The perpendicular magnetic field generated by electric current flowing into the stripe has a gradient in the transverse direction. Therefore it may cause a deflection of the skyrmion, and compete with the deflection caused by gyrotropic effects.

\begin{figure}[h!]
\includegraphics[width=12cm]{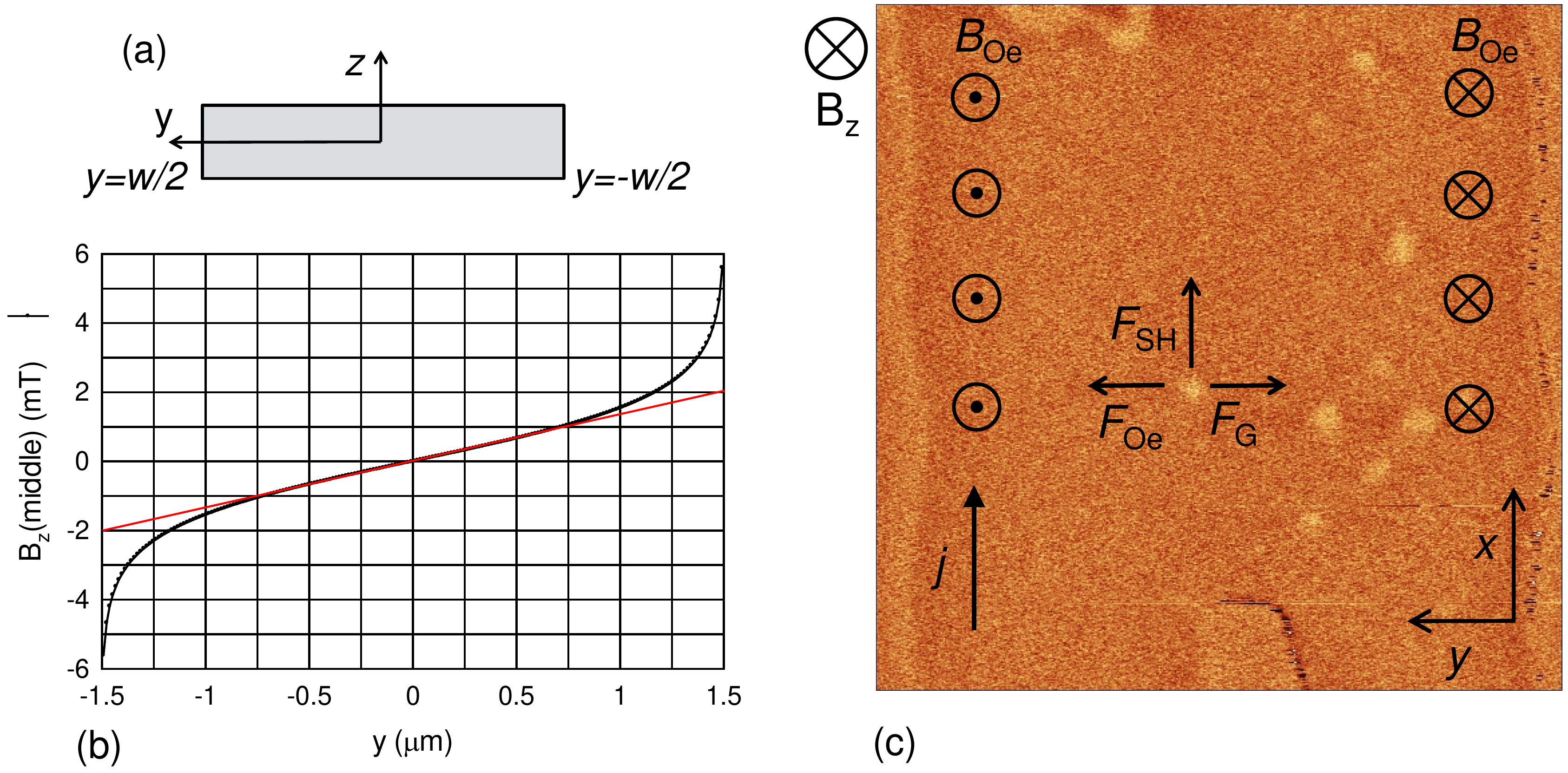}
\caption{\textbf{Effect of the Oersted field on the skyrmion deflection.} (a) Sketch of the stripe, aligned along $x$, for the Oersted field calculation. (b) Variation of the $z$ component of the Oersted field along the transverse direction to the stripe for $j=3\times10^{11}$~A/m$^2$. (c) Superposed on the experimental result from Fig.~3(b), forces acting on a skyrmion with a positive core polarisation $p=1$. The current $j$ along $x$ pushes the skyrmion along the stripe through the spin Hall effect (related force: $F_\mathrm{SH}$) which acquires a velocity $v$ in the same direction. The current also produces an Oersted field (positive at the left stripe edge) which attracts the skyrmion to the left edge (related force: $F_\mathrm{Oe}$). Due to the velocity and its particular topology, the skyrmion experiences a gyrotropic force, which attracts the skyrmion toward the right edge (related force: $F_\mathrm{G}$). The experimental deflection toward the right edge proves the dominant role of the gyrotropic force with respect to the Oersted field-induced deflection. \label{FigOe}}
\end{figure}

In our situation, the magnetic film is surrounded by conducting layers so that the Oersted field is essentially perpendicular to the stripe plane, and only depends on the transverse coordinate $y$:
\begin{equation}
B_z(y)=\frac{\mu_0 j h}{4 \pi} \left[\ln\left(\frac{(h/2)^2+(w/2+y)^2}{(h/2)^2+(w/2-y)^2}\right)+8\frac{w/2+y}{h/2}\arctan\left( \frac{h/2}{w/2+y}\right)-8\frac{w/2-y}{h/2}\arctan\left( \frac{h/2}{w/2-y}\right) \right]
\end{equation}
with $h$ and $w$ respectively the metallic stripe thickness and width (see Fig.~\ref{FigOe}(a)). Note that the magnetic sample thickness $t=3$~nm differs from $h=16$~nm, which includes non-magnetic layers.
A skyrmion at the center of the stripe will tend to minimize its energy by shifting towards the edge where the field is parallel to its core magnetization orientation, thus in a direction parallel to $p j \vec{y}$.

Under the same current, the skyrmion moves along the current direction and therefore feels a gyrotropic force. For the initial motion, i.e. purely along $\vec{x}$, this force is along $-pv_x \vec{y}\propto-pj \vec{y}$.

It is remarkable to note that these two forces have the same dependence on both $j$ and $p$ and are always antiparallel (see Fig.~\ref{FigOe}(c)). Note that this property is valid only in our specific case where skyrmions move along the current flow direction. Therefore, as our experiments indicate a deflection which is coherent with the direction predicted by the gyrotropic effect, there is no ambiguity on the dominant mechanism.

At the center of the stripe, the corresponding forces can be quantitatively evaluated. In a large region in the stripe center, the Oersted field varies linearly with $y$ as
\begin{equation}
	B_{z}\cong \frac{\mu_0 j h}{4 \pi} \frac{8y}{w}.
\end{equation}
Deriving the energy of a skyrmion of radius $R$ with respect to $y$ gives the Oersted field force
\begin{equation}
	\vec{F}_\mathrm{Oe}=2 M\mathrm{_s} \frac{\partial B_z}{\partial y} \pi R^2 t y= 4 \mu_0 M_s j \frac{h R^2 t}{w} \vec{y}.
\end{equation}
The gyrotropic force is
\begin{equation}
	\vec{F}_\mathrm{G}=-\frac{4\pi\mu_0M_\mathrm{s}t}{\gamma_0}v_x \vec{y}
\end{equation}
and the force exerted by the spin Hall effect is
\begin{equation}
\vec{F}_{\mathrm{SH}}\approx\frac{\hbar }{2 e} \theta_{\mathrm{SH}} j \pi^2  R \vec{x}
\end{equation}
which is valid for a large skyrmion diameter as compared to the DW width.

Considering the parameters of our experiment ($M_\mathrm{s}=850$~kA/m, $t=3$~nm, $\theta_{\mathrm{SH}}\approx0.1$~[\onlinecite{nguyen2016spin}], $h=16$~nm and $w=1$ or 3~$\mu$m) we can estimate these forces. At our maximum current density $j=4.4 \times 10^{11}$~A/m$^2$, the skyrmion velocity is about $v_x\approx60$~m/s, which gives values summarized in Table I where the two estimations of $F_\mathrm{Oe}$ correspond to the two experimental setups shown in Figs.~2 and 3 respectively.

\begin{table}[h!]

    \begin{minipage}{.4\linewidth}
\begin{tabular}{c|c|c|c|c}
  $R$ & 60~nm & 80~nm & 100~nm & 150~nm  \\ \hline
  $\lvert F_\mathrm{SH}\rvert$ & 8.6 & 11.0 & 14.0 & 21.4 \\
  $\lvert F_\mathrm{G}\rvert$ & 11.0 & 11.0 & 11.0 & 11.0 \\
  $\lvert F_\mathrm{Oe}\rvert$ & 0.3 & 0.6 & 0.9 & 2.0 \\
\end{tabular}  \\
(a)
\end{minipage}%
    \begin{minipage}{.4\linewidth}
\begin{tabular}{c|c|c|c|c}
  $R$ & 60~nm & 80~nm & 100~nm & 150~nm  \\ \hline
  $\lvert F_\mathrm{SH}\rvert$ & 8.6 & 11.0 & 14.0 & 21.4 \\
  $\lvert F_\mathrm{G}\rvert$ & 11.0 & 11.0 & 11.0 & 11.0 \\
  $\lvert F_\mathrm{Oe}\rvert$ & 0.1 & 0.2 & 0.3 & 0.7 \\
\end{tabular}\\
(b)
\end{minipage}%
\caption{Forces acting on a moving skyrmion in (a) 1~$\mu$m and (b) 3~$\mu$m wide wires. Values are given in pN.}
\end{table}

This quantitative estimation proves again that the gyrotropic deflection dominates over the Oersted field-induced deflection.

\section{Skyrmion motion under spin Hall effect in a perfect medium}

In general, two very different skyrmions motion regimes are known \cite{iwasaki2013universal, sampaio2013nucleation}. For a free-standing skyrmion \cite{iwasaki2013universal}, the solution of the Thiele equation gives a  velocity having both longitudinal and transverse components
\begin{eqnarray}\label{velocity1}
v_x=\frac{\alpha \mathcal{D}_{xx} F_{\mathrm{SH}} - GF_{\mathrm{Oe}}}{(\alpha \mathcal{D}_{xx})^2+G^2},\\ \label{velocity2}
v_y=\frac{\alpha \mathcal{D}_{xx} F_{\mathrm{Oe}} + G F_{\mathrm{SH}}}{(\alpha \mathcal{D}_{xx})^2+G^2}.
\end{eqnarray}
with $\alpha$ the Gilbert damping, $\mathcal{D}_{xx}$ the diagonal element of the dissipative tensor and $G$ the $z$ component of the gyrovector. We note that one has to take into account the absolute sign of the acting forces in these equations.

The second mechanism has been proposed by Sampaio \textit{et al.} \cite{sampaio2013nucleation}. In a confined geometry where the lateral forces are balanced by edge repulsion, the skyrmion velocity increases significantly to
\begin{equation}\label{Eq_Sampaio}
v_x=\frac{F_{\mathrm{SH}}}{\alpha \mathcal{D}_{xx}}.
\end{equation}

To quantitatively express the forces acting on the moving skyrmion, we need to estimate $G$ and $\mathcal{D}_{xx}$.
On the one hand, from the interpretation of the gyrovector as the surface covered by the magnetic texture on the unit sphere \cite{thiele1973steady}, we know that the gyrovector value is, independently of the skyrmion profile,
\begin{equation}
G=-4 \pi\frac{\mu_0M_\mathrm{s}t}{\gamma_0}.
\end{equation}
On the other hand, the value of the dissipation depends on the skyrmion profile.
It is expressed as
\begin{equation}
\mathcal{D}_{xx}= \frac{\mu_0M_\mathrm{s}t}{\gamma_0} \int{\left(\frac{\partial \vec{m}}{\partial x}\right)^2 \mathrm{d}x\mathrm{d}y}=  \frac{\pi\mu_0M_\mathrm{s}t}{\gamma_0} \int{\left[ \left(\frac{d\theta}{\mathrm{d}r}\right)^2 + \frac{\sin^2\theta}{r^2} \right] r\mathrm{d}r},
\label{eq:Dxx}
\end{equation}
the latter equality applying to a profile with revolution symmetry.

\begin{figure}[h!]
\includegraphics[width=17cm]{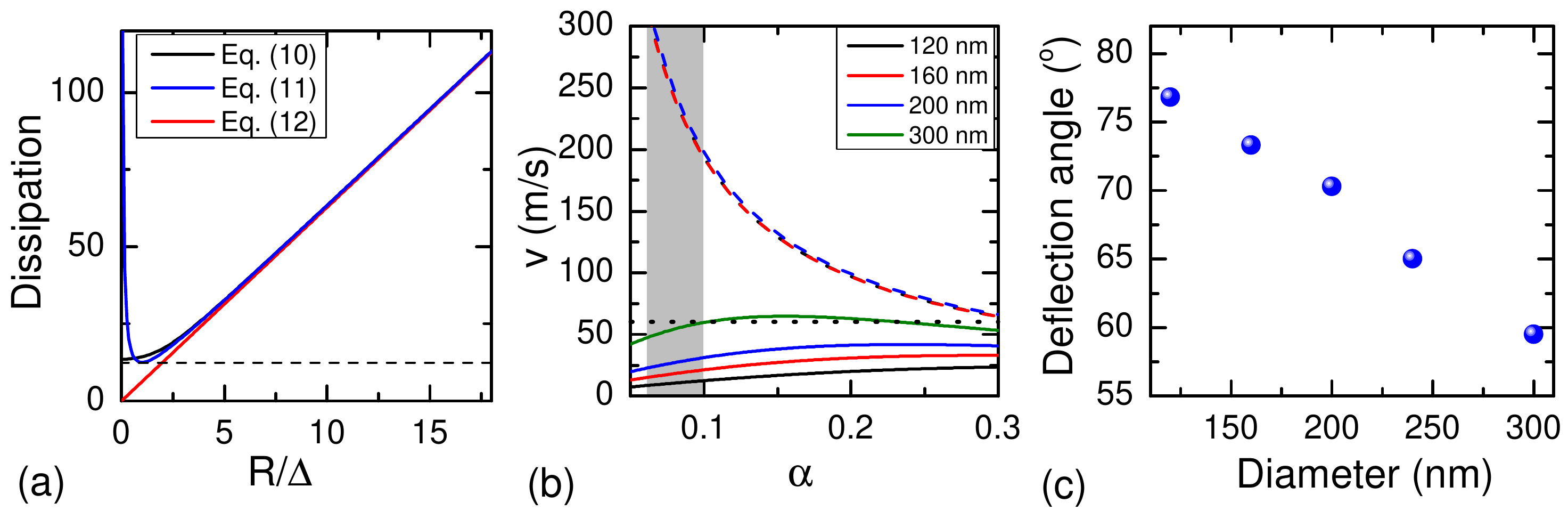}
\caption{\textbf{Skyrmion dynamics.} (a) Calculated diagonal element of the dissipative tensor using Eqs. (\ref{eq:Dxx}), (\ref{eq:Dxx2}) and (\ref{eq:Dxx1}) normalized by the prefactor $\frac{\mu_0M_\mathrm{s}t}{\gamma_0}$. The Belavin-Polyakov limit\cite{belavin1975} ($4\pi$) is plotted in dashed line. (b) Calculated skyrmion velocities using equations (\ref{velocity1}) (lines) and (\ref{Eq_Sampaio}) (dashed) in case of $w=1~\mu$m$, \theta_{\mathrm{SH}}$=0.1 and $j=0.44$~TA/m$^2$ for skyrmions of various diameters. The dotted line corresponds to the experimental value. (c) Deflection angle at $j=0.44$~TA/m$^2$ and $\alpha=0.08$ for various skyrmion sizes.
\label{Fig2}}
\end{figure}

The numerical evaluation of $\frac{\gamma_0}{\mu_0M_\mathrm{s}t}\mathcal{D}_{xx}$ using the planar 360$^{\circ}$ Bloch DW profile, that was shown to fit well the skyrmion radial profiles as observed by spin-polarized STM \cite{romming2015field}, is shown in Fig.~\ref{Fig2}(a), as a function of the reduced ratio $R/\Delta$.
This result is compared to two approximations of the radial integral in Eq.~(\ref{eq:Dxx}), relevant when one has $R \gg \Delta$.
In such a case one can put $r=R$ when integrating over the DW profile, and assume that $\theta(r)$ has the Bloch wall profile.
The solution leads to
\begin{equation}
\mathcal{D}_{xx} \approx \frac{\mu_0M_\mathrm{s}t}{\gamma_0}  2\pi \left(\frac{R}{\Delta}+\frac{\Delta}{R}\right),
\label{eq:Dxx2}
\end{equation}
when one considers both terms, whereas with only the first term one gets
\begin{equation}
\mathcal{D}_{xx} \approx \frac{\mu_0M_\mathrm{s}t}{\gamma_0} 2\pi \frac{R}{\Delta}.
\label{eq:Dxx1}
\end{equation}
One sees that for $R > \Delta$ the approximation of Eq.~(\ref{eq:Dxx2}) corresponds well to the result for the 360$^{\circ}$ DW profile.
The dissipation values are distinctly smaller than those quoted in Ref.~\onlinecite{jiang2016direct}, in which a very schematic DW profile was used to compute the first term of radial integral of Eq.~(\ref{eq:Dxx}).
Note that, even if the dissipation depends on the skyrmion profile, it cannot take any value.
Indeed, Eq.~(\ref{eq:Dxx}) shows that the trace of the dissipation tensor ($\mathcal{D}_{xx}+\mathcal{D}_{yy}$) is proportional to the exchange energy.
At this point, it is useful to recall that Belavin and Polyakov have proved mathematically\cite{belavin1975} that the exchange energy is bounded by the absolute value of the skyrmion number, the equality being obtained for the Belavin-Polyakov profile.
This means, in the present notations, that
\begin{equation}
\mathcal{D}_{xx}+\mathcal{D}_{yy} > 2 |G|.
\label{eq:BP}
\end{equation}
Thus for a profile with revolution symmetry where $\mathcal{D}_{xx}=\mathcal{D}_{yy}$, and a skyrmion number $\pm 1$ one has, mathematically,
\begin{equation}
\mathcal{D}_{xx} > 4\pi \frac{\mu_0M_\mathrm{s}t}{\gamma_0}.
\end{equation}

After this digression on the evaluation of the dissipation tensor of a skyrmion, we come back to numerical estimates.
With $R=80$~nm and $\Delta=12.9$~nm (see Fig.~1(b) of the main text; note that this value is very different from the 1-layer calculation $\sqrt{A/K_\mathrm{eff}}$ due to dipolar effects), we obtain $\mathcal{D}_{xx}\gamma_0/\mu_0 M_\mathrm{s} t\approx 40$. To express the effect of $\alpha$ on the velocity, the calculated velocities are plotted in Fig.~\ref{Fig2}(b).

In the case of the first regime the velocity at $j=4.4\times10^{11}$ A/m$^2$ along $x$ is expected to be 18~m/s resulting in a deflection angle of $\approx73^{\circ}$ (see Fig.~\ref{Fig2}(b)), while for the second regime, a velocity of 240~m/s is expected (Note that, in this case, the velocity strongly depends on $\alpha$, for which our rough estimation implies a large uncertainty on the expected velocity.). The experimental value (60~m/s) is close to the free-standing skyrmion solution. The fact that the calculated skyrmion velocities in the confined skyrmion regime are independent of the skyrmion size arises from the proportionality of $F_\mathrm{SH}$ and $\mathcal{D}_{xx}$ to $R$ for $R>\Delta$.


\end{document}